\documentstyle[floats,aps,preprint]{revtex}
\begin{document}
\draft 
\preprint{\vbox{{\hbox{SOGANG-HEP 280/01}}}}
\title{Entropy of the Randall-Sundrum black brane world
        in the brick-wall method}
\author
{Won Tae Kim\footnote{electronic address:wtkim@ccs.sogang.ac.kr},
John J. Oh\footnote{electronic address:john5@string.sogang.ac.kr},
and Young-Jai Park\footnote{electronic address:yjpark@ccs.sogang.ac.kr}}
\address{Department of Physics and Basic Science Research Institute,\\
         Sogang University, C.P.O. Box 1142, Seoul 100-611, Korea}
\date{\today}
\maketitle
\bigskip
\begin{abstract}
We calculate the entropy of the brane-world black hole in the
Randall-Sundrum(RS) model by using the brick-wall method. The modes
along the extra dimension are semi-classically quantized on the extra
dimension. The number of modes in the extra dimension is given as a
simple form with the help of the RS mass relation,
and then the entropy for the scalar modes in the five-dimensional spacetime
is described by the two-dimensional area of the black brane world.
\end{abstract}
\bigskip

\newpage
Recently, there has been much interests in
the Randall and Sundrum(RS) model to resolve the gauge hierarchy problem
\cite{rs1,rs2}, which is based upon the fact that our universe may be
embedded in higher-dimensional spacetimes
\cite{add,aadd}. Furthermore, the various aspects of this model have
been studied on the cosmological grounds \cite{cos}. 
In order to study the gauge hierarchy problem, RS have proposed a
two-brane model called RS1 model involving
a small and curved extra dimension, which is a slice of anti-de Sitter
(AdS) spacetime~\cite{rs1}. The negative tension brane is regarded as
our universe, and the hierarchy between physical scales naturally
appears in our brane. Furthermore, they have studied a single-brane
model(RS2 model) by taking $r_{c} \rightarrow \infty$, where $r_{c}$
is a radius of the extra dimension~\cite{rs2}. In these models, the
nonfactorizable metric is essential as a static anti-de Sitter(AdS) spacetime,
which is different from that of the
conventional Kaluza-Klein(KK) style in that the extra coordinate is
associated with a conformal factor.
On the other hand, the static AdS domain wall has been already found as
BPS domain walls of supergravity theories in
Ref. \cite{cgr}, which is relevant to the RS model in a special case.

Subsequently, it has been shown that the curved domain wall solution
in the RS models can be given
by transforming the Minkowskian metric by the Schwarzschild one,
which is described by a black string solution in
five dimensions \cite{chr}. The $d$-dimensional generalization of
this has been shown in Ref. \cite{kooy} and a brane-world black hole
for a rotating case has recently been studied in Ref. \cite{mps}. In
Refs. \cite{ehm,ehm2}, thermodynamics and aspects of evaporation in
brane-world black holes have been studied.
To study quantum mechanical aspect of this black brane world (black
string intersecting the brane world), we
may first consider its entropy, which is expected
to satisfy the area law \cite{haw}. In this paper, we would like to
calculate the statistical entropy of the black brane world in terms of the brick wall
method \cite{tho,dlm,ff,kkps}.
We first derive the brane-world black hole solution directly starting from
the equation of motion, and obtain the black brane world system. And then we semi-classically quantize the massive scalar
field on this five-dimensional background. First, the modes along the extra
dimension is quantized, and then the mass is naturally discretized.
Next, we calculate the other modes following the brick
wall method, and obtain the desired entropy of the brane-world black
hole by using the RS relation. This shows that the number of modes in the
five-dimensional spacetime are can be effectively described by the
two-dimensional surface. 

Let us now consider the RS model in $(4+1)$ dimensions,
\begin{equation}
 \label{2baction}
 S_{(5)} = \frac{1}{2 \kappa^2_{(5)}} \int d^4 x \int dy \sqrt{-g_{(5)}}
            \left[ R^{(5)} + 12k^2 \right]
         - \int d^4 x \left[\sqrt{-g_{(+)}} \lambda_{(+)} +
\sqrt{-g_{(-)}} \lambda_{(-)} \right],
\end{equation}
where $\lambda_{(+)}$ and $\lambda_{(-)}$ are tensions of the branes
at $y=0$ and $y=\pi r_c$, respectively, $12k^2$ is a cosmological
constant, and $\kappa_{(5)}^2=8\pi G_{N}^{(5)}$. 
Note that we use $M,N,K, \cdots = 0, 1, \cdots, 4$ for
$(4+1)$-dimensional spacetime indices and $\mu, \nu, \kappa, \cdots =
0, 1, \cdots, 3$ for branes. We assume orbifold $S^1/Z_{2}$ which has
a periodicity in the extra coordinate $y$, and identify $-y$ with
$y$. Two singular points on the orbifold are located at $y=0$ and
$y=\pi r_c$, and two 3-branes are placed at these points,
respectively. Note that the metric on each brane is defined as $
g_{\mu \nu}^{(+)} \equiv g_{\mu\nu}^{(5)}(x^{\mu},y=0)$ and $g_{\mu
  \nu}^{(-)} \equiv g_{\mu\nu}^{(5)}(x^{\mu}, y=\pi r_c)$, respectively.

We now assume the bulk  metric as
\begin{equation}
   \label{metrican}
  ds^2_{(5)}=e^{-2k|y|\Phi(x)} g_{\mu\nu}(x) dx^{\mu}dx^{\nu} + T^2(x)dy^2,
\end{equation}
where the moduli field $T(x)$ is different from $\Phi (x)$ for the
present. From Eq.~(\ref{2baction}),
the equation of motion is given as
\begin{equation}
\label{einseqnss}
     G_{MN}^{(5)} = T_{MN}^{(5)}.
\end{equation}
By using the metric (\ref{metrican}), the Einstein tensors are calculated as
\begin{eqnarray}
 G_{\mu\nu}^{(5)} &=& G_{\mu\nu} - \frac{1}{T} \left(\nabla_{\mu}
                       \nabla_{\nu} T - g_{\mu\nu} \Box T \right)
                       +2k|y| \left(\nabla_{\mu}\nabla_{\nu} \Phi -
                       g_{\mu\nu} \Box \Phi \right) \nonumber \\
                  & & - \frac{k|y|}{T} \left(\nabla_{\mu}T
                       \nabla_{\nu}\Phi + \nabla_{\mu}\Phi
                       \nabla_{\nu}T + g_{\mu\nu}
                       \nabla_{\rho}T\nabla^{\rho} \Phi \right) \nonumber \\
                  & & + k^2 |y|^2 \left( 2 \nabla_{\mu}\Phi
                       \nabla_{\nu}\Phi + g_{\mu\nu} (\nabla
                       \Phi)^2 \right) + \frac{6k\Phi}{T^2} e^{-2k|y|
                       \Phi} g_{\mu\nu} \left( k\Phi - \delta(y) +
                       \delta(y - \pi r_{c}) \right),
  \label{einseq1} \\
 G_{\mu y}^{(5)}  &=& 3k \left( \partial_{\mu} \Phi -
                   \frac{\Phi}{T} \partial_{\mu}T \right) \partial_{y} |y|,
  \label{einseq2} \\
 G_{yy}^{(5)}     &=& - \frac{1}{2} T^2 e^{2k|y| \Phi} \left( R +6k|y|
                  \Box \Phi - 6k^2 |y|^2 (\nabla \Phi)^2 \right.
                 - \left. \frac{12k^2 \Phi^2}{T^2} e^{ -2k|y| \Phi} \right),
  \label{einseq3}
\end{eqnarray}
and the stress-energy tensor is explicitly written as
\begin{eqnarray}
 T_{MN}^{(5)} = 6 k^2 g_{MN}^{(5)} &+& \kappa_{(5)}^2
                 \frac{\sqrt{-g_{(+)}}}{\sqrt{-g_{(5)}}}
                 \lambda_{(+)} \delta(y) g_{\mu\nu}^{(+)}
                 \delta_{M}^{\mu} \delta_{N}^{\nu} \nonumber \\
             &+& \kappa_{(5)}^2 \frac{\sqrt{-g_{(-)}}}{\sqrt{-g_{(5)}}}
                 \lambda_{(-)} \delta(y-\pi r_c)
                 g_{\mu\nu}^{(-)} \delta_{M}^{\mu} \delta_{N}^{\nu}.
 \label{stress}
\end{eqnarray}
Since ($\mu y$)-component of Eq. (\ref{stress}) vanishes, from
Eqs. (\ref{einseqnss}) and (\ref{einseq2}) we obtain the following relation,
\begin{equation}
 G_{\mu y}^{(5)} = 3k \left( \partial_{\mu} \Phi -
 \frac{\Phi}{T} \partial_{\mu}T\right) \partial_{y} |y| = 0,
 \label{consteq}
\end{equation}
which yields simply $ \Phi(x) = T(x)$.

In the ($\mu\nu$)-component of Eq. (\ref{einseqnss}), there exist
discontinuities resulting from the delta functional source due to the
presence of brane tensions at $y=0$ and $y=\pi r_c$. At this stage,
we now consider junction conditions \cite{junction}, and integrate out
the Einstein equation near the branes,
\begin{eqnarray}
  & &\int_{0-\epsilon}^{0+\epsilon}dy G_{\mu\nu}^{(5)}
     = \int_{0-\epsilon}^{0+\epsilon} dy T_{\mu\nu}^{(5)}, \nonumber \\
  & &\int_{\pi r_c-\epsilon}^{\pi r_c+\epsilon}dy G_{\mu\nu}^{(5)}
     = \int_{\pi r_c-\epsilon}^{\pi r_c+\epsilon}dy T_{\mu\nu}^{(5)}.
 \label{integrated}
\end{eqnarray}
The jump along the extra coordinate near the 3-branes gives a relation
between brane tensions and the cosmological constant,
\begin{equation}
 \label{rel2b}
  \lambda_{(+)} = - \lambda_{(-)} = \frac{6}{\kappa_{(5)}^2} k,
\end{equation}
where we note that the branes at $y=0$ and $y=\pi r_c$ have a
positive tension ($\lambda_{(+)}$) and a negative one
($\lambda_{(-)}$), respectively. Using the relation (\ref{rel2b}) for
this brane model, the equation of motion (\ref{einseqnss}) is
explicitly given as
\begin{eqnarray}
 & & R_{\mu \nu} - \frac12 g_{\mu\nu} R - \frac{1}{T} \left[
       \nabla_{\mu} \nabla_{\nu} T - g_{\mu\nu} \Box T\right] + k^2
       y^2 \left[  2\nabla_{\mu}T\nabla_{\nu}T + g_{\mu\nu} (\nabla
       T)^2  \right] \nonumber \\
 & & \qquad+ 2k|y|\left[ \nabla_{\mu} \nabla_{\nu} T - g_{\mu\nu} \Box
       T\right] - \frac{k|y|}{T}\left[2 \nabla_{\mu}T\nabla_{\nu}T +
       g_{\mu\nu} (\nabla T)^2 \right]=0, \label{eqnmot1} \\
 & & R + 6k|y| \Box T - 6k^2 y^2 (\nabla T)^2 = 0. \label{eqnmot2}
\end{eqnarray}
Traces of Eq. (\ref{eqnmot1}) and Eq. (\ref{eqnmot2}) give the
following reduced equations,
\begin{eqnarray}
  R + 3k|y| \Box T = 0, \nonumber \\
  \Box T - 2k|y| (\nabla T)^2 = 0.
 \label{reduceqn}
\end{eqnarray}
In Eq.~(\ref{reduceqn}), as a simple constant solution of $T(x)$, let
us set $T(x)=1$. Then metric solution $g_{\mu\nu}$ should be determined
by $ R = 0$. Combining $R = 0$ and $T(x)=1$ with Eq. (\ref{eqnmot1}), the Ricci
flat condition, $R_{\mu\nu} = 0$, introduced in Ref. \cite{chr}, is
obtained from the equations of motion. From this condition, it is
natural to consider the 4-dimensional Schwarzschild black hole
solution as a slice of AdS spacetime as a brane solution,
\begin{equation}
  ds^2 = e^{-2k|y|}\left[ - \left(1-\frac{2M}{r}\right) dt^2
        + \left(1-\frac{2M}{r}\right)^{-1} dr^2
        + r^2 d\Omega^2_{(2)} \right] + dy^2,
 \label{d4ss}
\end{equation}
where $d\Omega^2_{(2)}$ is a metric of unit 2-sphere and we set
$G_{(4)} = 1$ for convenience. It is a black string solution
intersecting the brane world, which
describes a black hole placed on the hypersurface at the fixed extra
coordinate. Arnowitt-Deser-Misner(ADM) mass $\tilde{M}$ of
the brane-world black hole measured on the brane is
$ \tilde{M} = M e^{-ky_{0}}$ where $y_{0}$ is $0$ or $\pi
r_{c}$. Then, ADM mass would be exponentially suppressed as $\tilde{M}
= M e^{-ky_{c}} |_{y_{c}\rightarrow\infty} \rightarrow 0$ on the
negative tension brane for the RS2 model. In our work, we shall focus
on the black hole placed at $y=0$.

Let us now study the black hole entropy in terms of the
brick wall method. 
With the help of the RS relation which connects the bulk mass 
with the mass on
the brane, the quantization will be simplified. We now assume
the massive scalar field $f$ on the five-dimensional background (\ref{d4ss}),
\begin{equation}
  \label{wveqn}
  (\Box_{(5)} - m^2)f = 0,
\end{equation}
and it is explicitly given as
\begin{eqnarray}
  \label{wveqn2}
  e^{2k|y|}& & \left[ -\frac{1}{g} \partial_{t}^2 f + \frac{1}{r^2}
  \partial_{r}\left(r^2 g \partial_{r} f\right) + \frac{1}{r^2 {\rm
  sin}\theta } \partial_{\theta} ({\rm sin} \theta \partial_{\theta}
  f) + \frac{1}{r^2 {\rm sin}^2 \theta} \partial_{\phi}^2 f\right]
  \nonumber \\
  & & + e^{4k|y|} \partial_{y}(e^{-4k|y|} \partial_{y} f) - m^2 f = 0,
\end{eqnarray}
where $g = g(r) = 1- \frac{2M}{r}$.
If we set
\begin{equation}
\label{5d}
 e^{4k|y|}  \partial_{y}(e^{-4k|y|} \partial_{y} \chi) -
m^2 \chi  + \mu^2 e^{2k|y|}\chi = 0,
\end{equation}
where $ f(t,r,\theta, \phi, y) \equiv \Psi(t,r,\theta, \phi)\chi(y)$,
then the separation of variables is easily done, and
the reduced form of the effective field equation becomes
\begin{equation}
  \label{eqn2}
   -\frac{1}{g} \partial_{t}^2 \Psi + \frac{1}{r^2} \partial_{r} \left(r^2g \partial_{r} \Psi\right) + \frac{1}{r^2 {\rm sin} \theta} \partial_{\theta} ({\rm sin}\theta \partial_{\theta} \Psi) + \frac{1}{r^2 {\rm sin}^2 \theta} \partial_{\phi}^2 \Psi - \mu^2 \Psi = 0.
\end{equation}
Note that the above eigenvalue $\mu^2$ plays a role of the effective
mass on the brane. Then further separation of variables 
is straightforward, which gives
\begin{equation}
  \label{eqn3}
  \frac{1}{r^2}(r^2 g \partial_{r} R) + \left( \frac{\omega^2}{g} - \frac{\ell (\ell + 1)}{r^2} - \mu^2 \right) R = 0,
\end{equation}
where $\Psi(t,r,\theta,\phi)\equiv R(r)S(\theta)e^{i\alpha \phi -
  i\omega t}$.
The exact quantization of Eq.(\ref{5d}) seems to be cumbersome.
However, in the WKB approximation \cite{tho}, the wave number $k_{\chi}$
of the wave function $\chi(y)$ is easily written as
\begin{equation}
  \label{waveno1}
  k_{\chi}^2(y, \mu) = \mu^2 e^{2k|y|} - m^2.
\end{equation}
Therefore, the number of modes $n_{\chi}$ is obtained as
\begin{eqnarray}
  \label{modes1}
 \pi n_{\chi}(\mu) &=& \int_{0}^{\pi r_{c}} dy k_{\chi}(y,\mu) \nonumber \\
                   &=& - \frac{\sqrt{\mu^2-m^2}}{k} + \frac{m}{k}
                   {\rm tan}^{-1}\left[\frac{\sqrt{\mu^2 -
                   m^2}}{m}\right] \nonumber \\
                   &+& \frac{\sqrt{\mu^2 e^{2k\pi r_{c}}-m^2}}{k} -
                   \frac{m}{k} {\rm tan}^{-1}\left[\frac{\sqrt{\mu^2
                   e^{2k\pi r_{c}}-m^2}}{m}\right].
\end{eqnarray}
Unfortunately, it is difficult to write the eigenvalue $\mu$
as a function of $n_{\chi}$ since the inverse function is
not easily obtained.
However, we now consider the RS relation
\cite{rs1}, which is key to the breakthrough.
It has been used
to connect the effective four-dimensional
mass with the five-dimensional bulk mass, and
it is given by $\mu =m e^{-ky_c}$ at the branes. We note
that in Ref. \cite{rs1} to resolve the hierarchy problem
between physical couplings, the negative tension brane at $y_{c} = \pi
r_{c}$ is regarded as our universe. 
On the other hand, in our black hole case,
the positive tension brane has been taken as our spacetime
of black brane world since the negative tension brane has naked
singularities. Furthermore, at the negative tension brane
the real wave number can not be defined.
So, for the present case, it is possible to take $y_c=0$
without these problems. Then the RS relation
is reduced to
\begin{equation}
  \label{RSrel}
 \mu = m,
\end{equation}
which yields
\begin{equation}
\label{modeschi}
\mu = \mu_{n_\chi}=\frac{\pi k}{\gamma} n_{\chi} ~~~~~~(n_\chi =1,2,3, \cdots)
\end{equation}
from Eq. (\ref{modes1}) with $\gamma = \sqrt{e^{2k\pi r_{c}} - 1} -
{\rm tan}^{-1} \sqrt{e^{2k\pi r_{c}} - 1}$.
If we take $r_c \rightarrow \infty$, then the mode spectrum is continuous 
similarly to the quantization on a circle. 

At last, the radial wave number
$ k_{R}^2 = g^{-2}\left[ \omega^2 - g \left(\frac{\ell (\ell +
      1)}{r^2} + \mu^2\right)\right]$ from Eq. (\ref{eqn3}) is semi-classically quantized as
\begin{equation}
  \label{quant}
  \pi n_{R}(\omega,\ell, \mu) = \int_{2M + h}^{L} dr g^{-1} \sqrt{\omega^2 - g\left( \frac{\ell (\ell+1)}{r^2} + \mu^2\right)},
\end{equation}
where $h$ and $L$ are ultra and infrared cutoffs, respectively which
are needed in the brick wall formalism.
Now the degeneracy for a given energy is defined by
\begin{eqnarray}
  \label{totalN}
  g(\omega) \equiv \pi N &=& \int d\ell (2 \ell + 1) \int dn_{\chi}
  \pi
n_{R}(\omega, \ell, n_{\chi}) \nonumber \\
&=& \int d\ell (2\ell +1) \int dn_{\chi} \int_{2M+h}^{L} dr g^{-1} \sqrt{ \omega^2 - g \left( \frac{\ell (\ell+1)}{r^2} + \lambda^2 n_{\chi}^2\right)},
\end{eqnarray}
where $\lambda = k \pi \gamma^{-1}$.

The free energy $F$ for this black brane world system is given as
\begin{equation}
  \label{freeenergy}
  e^{-\beta F} = \sum e^{-\beta \omega} = \prod_{n_{R}, \ell, n_{\chi}} \frac{1}{1- e^{-\beta \omega}},
\end{equation}
which is explicitly written as
\begin{eqnarray}
  \label{freeenrgy}
  \pi \beta F &=& - \int_{0}^{\infty} d\omega \frac{\beta
 g(\omega)}{e^{\beta \omega} - 1} \nonumber \\
 &=&  - \beta \int_{0}^{\infty} d\omega (e^{\beta \omega} -1)^{-1} \int d\ell (2\ell +1) \int dn_{\chi} \int_{2M+h}^{L} g^{-1} \sqrt{ \omega^2 - g \left( \frac{\ell (\ell+1)}{r^2} + \lambda^2 n_{\chi}^2\right)}.
\end{eqnarray}
For the reality condition of the free energy,
the integration ranges are restricted to 
$  0 \le n_{\chi} \le \left(\sqrt{\omega^2/g - \ell (\ell + 1)/r^2}\right)/{\lambda}$ and
$  0 \le \ell \le \left(-1+\sqrt{1+ {4 \omega^2 r^2}/g}\right)/2$.

In the approximation of $L>>2M$,
the free energy is evaluated as
\begin{equation}
 \label{free}
F \approx  - 64 \frac{\gamma \zeta(5)}{\pi h^2} \sqrt{\frac{h}{2M}} \left(\frac{M}{\beta}\right)^5 - \frac{3}{8\pi k} \gamma L^3 \int_{0}^{\infty} d\omega \frac{\omega^4}{e^{\beta \omega} - 1}.
\end{equation}
It is interesting to note that Eq. (\ref{free}) is more or less
different from the four-dimensional free energy calculation
in the Schwarzschild black hole background \cite{tho}, which is given as
\begin{equation}
  \label{free4d}
  F \approx - \frac{2\pi^3}{45 h} \left(\frac{2M}{\beta}\right)^4 - \frac{2}{9 \pi}
  L^3 \int_{m}^{\infty} d\omega \frac{(\omega^2 - m^2)^{3/2}}{e^{\beta
  \omega} - 1}.
\end{equation}
The reason why the mass term of the scalar field in the bulk spacetime 
does not appear in the free energy expression (\ref{free}) on the
contrary to the four-dimensional case is that 
the mass $m$ in the bulk can be effectively interpreted as that of 
the four-dimensional mass $\mu$ in terms of the RS relation
(\ref{RSrel}), and it is integrated out in the free-energy calculation after
semi-classical quantization along the extra dimension.
 
Now the Hawking temperature is defined by 
\begin{equation}
  \label{hawk}
  T_{H} = \frac{\kappa_{H}}{2\pi}
\end{equation}
where $ \kappa_{H}^2 = -\frac{1}{2} (\nabla^{\mu} \chi^{\nu})(\nabla_{\mu}
  \chi_{\nu}) |_{r=r_{H}}$ and $\chi$ is a time-like Killing vector,
which yields 
\begin{equation}
  \label{hawking}
  T_{H} = \beta^{-1} = \frac{1}{8 \pi M}.
\end{equation}
Therefore, from the free energy (\ref{free}) the entropy is 
\begin{eqnarray}
  \label{entropy}
  S &=& \beta^2 \left(\frac{\partial F}{\partial
  \beta}\right)\nonumber \\
    &=& \frac{{\cal A}_{H}}{4 G_{(4)}} \nonumber \\
    &=& 4 \pi M^2,
\end{eqnarray}
as far as the brick wall is located at
$ h=\left(\frac{5 \gamma \zeta(5)}{2^{8}\pi^{6}} 
\right)^{2}\cdot \frac{1}{2M}$.
Of course, the invariance distance from the horizon is calculated so
that in the approximation, $M >> h$,
\begin{eqnarray}
  \int_{r=2M}^{r=2M+h} \frac{dr}{\sqrt{1-\frac{2M}{r}}} &\approx& \sqrt{2Mh}
  \nonumber \\
  &=& \left(\frac{5 \gamma \zeta(5)}{2^{8} \pi^{6}}\right),
\end{eqnarray}
which is independent of the black hole mass.

We have calculated the entropy of the massive scalar field
in the RS model by using the brick-wall method. In our model,
the positive tension brane as a black hole solution
was considered in order to avoid curvature singularities
whereas in the cosmological consideration
the negative tension brane has been identified as our universe
to resolve the gauge hierarchy problem. 

In our result,
it is reminiscent of the holographic principle \cite{holo} 
of black hole
physics, so the number of degrees of freedom can be
derived by the horizon area of the brane, although
this does not mean that
the massive modes live on the brane.
It would be interesting whether or not 
the area law of the black hole entropy  
can be derived from the bulk theory by using the other methods.  

A final comments are in order. 
At first glance, the massless limit does not exist in
Eq. (\ref{RSrel}). However, from the beginning equation (\ref{waveno1}),
one can easily take the massless limit of the scalar field, and
the quantized rule is given as
$\pi n_{\chi} (\mu) = \int_{0}^{\pi r_{c}} dy 
\mu e^{2ky} = \frac{\mu}{2k} \left( e^{2k\pi r_{c}} - 1 \right)
  \approx \frac{\mu}{k} \lambda_{0}$
where $\lambda_{0} = (e^{2k\pi r_{c}} - 1)/2$.
As a result, the quantized rule is not changed except
the coefficient, that is  $\mu = k \pi / \lambda_{0} n_{\chi}$.
Therefore, the result for the massive case is still valid. The
physical significance for the massless limit
is that in fact the expected entropy can be obtained
without recourse to the RS mass relation (\ref{RSrel}),
while for the massive case the relation is helpful
in the derivation of the entropy. At this stage, we do not know
physically whether the RS relation (\ref{RSrel}) is essential or
not in deriving the entropy. We hope this problem will be discussed in
elsewhere. 

\vspace{1cm}

{\bf Acknowledgments}\\
This work was supported by grant No. 2000-2-11100-002-5
from the Basic Research Program of the Korea Science and
Engineering Foundation. We would like to thank P. Jung for
exciting discussions.

%%%%%%%%%%%%%%%%%%%% References %%%%%%%%%%%%%%%%%%%%%%%%%


\begin{references}
\bibitem {rs1} L. Randall and R. Sundrum, {\it Large Mass Hierarchy
    from a Small Extra Dimension}, Phys. Rev. Lett. {\bf 83} (1999) 3370
    , {\tt  [hep-ph/9905221]}.
\bibitem {rs2} L. Randall and R. Sundrum, {\it An Alternative to
    Compactification}, Phys. Rev. Lett. {\bf 83} (1999) 4690,
    {\tt [hep-th/9906064]}.
\bibitem {add} N. Arkani-Hamed, S. Dimopoulos, and G. Dvali,
    {\it The Hierarchy Problem and New Dimensions at a Millimeter},
    Phys. Lett. {\bf B429} (1998) 263, {\tt [hep-ph/9803315]}.
\bibitem {aadd} I. Antoniadis, N. Arkani-Hamed, S. Dimopoulos, and
    G. Dvali, {\it New dimensions at a millimeter to a Fermi and
    Superstrings at a TeV}, Phys. Lett. {\bf B436} (1998) 257,
    {\tt [hep-ph/9804398]}.
\bibitem {cos} P. Bin{\'{e}}truy, C. Deffayet, and D. Langlois,
    {\it Non-conventional cosmology from a brane-universe},
    Nucl. Phys. {\bf B565} (2000) 269, {\tt [hep-th/9905012]};
    C. Cs{\'{a}}ki, M. Graesser, L. Randall, and
    J. Terning, {\it Cosmology of Brane Models with Radion
    Stabilization}, {\tt [hep-ph/9911406]}.
\bibitem {cgr} M. Cveti${\check{c}}$, S. Griffies, and S.-J. Rey, {\it
    Static Domain Walls in $N=1$ Supergravity}, Nucl. Phys. {\bf B381}
    (1992) 301 ; M. Cveti${\check{c}}$, S. Griffies, and H. Soleng, {\it
    Local and Global Gravitational Aspects of Domain Wall
    Space-times}, Phys. Rev. {\bf D48} (1993) 2613, {\tt
    [gr-qc/9306005]}.
\bibitem {chr} A. Chamblin, S. W. Hawking, and H. S. Reall,
    {\it Brane-World Black Holes}, Phys. Rev. {\bf D61} (2000) 065007,
    {\tt [hep-th/9909205]}.
\bibitem {kooy} W. T. Kim, J. J. Oh, M. K. Oh, and M. S. Yoon, {\it
    Brane-World Black Holes in Randall-Sundrum Models}, {\tt
    [hep-th/0006134]}.
\bibitem {mps} M. S. Modgil, S. Panda, and G. Sengupta, {\it Rotating
    Brane World Black Holes}, {\tt [hep-th/0104122]}.
\bibitem {ehm} R. Emparan, G. Horowitz, and R. C. Myers, {\it Exact
    Description of Black Holes on Branes}, J. of High Energy
    Phys. {\bf 001} (2000) 07, {\tt [hep-th/9911043]}.
\bibitem {ehm2} R. Emparan, G. Horowitz, and R. C. Myers, {\it Black
    Holes Radiate Mainly on the Brane}, Phys. Rev. Lett. {\bf 85}
    (2000) 499, {\tt [hep-th/0003118]}.
\bibitem{haw} J. D. Bekenstein, {\it Black Holes and Entropy},
    Phys. Rev. {\bf D7} (1973) 2333 ; S. W. Hawking, {\it Particle
    Creation by Black Holes}, Commun. Math. Phys. {\bf 43} (1975) 199.
\bibitem {tho} G. 't Hooft, {\it On the Quantum Structure of a Black
    Hole}, Nucl. Phys. {\bf B256} (1985) 727.
\bibitem {dlm} J.-G. Demers, R. Lafrance, and R. C. Myers, {\it Black hole
    entropy without brick walls}, Phys. Rev. {\bf D52} (1995) 2245 ;
    R. B. Mann, L. Tarasov, and A. Zelnikov, {\it Brick walls for
    black holes}, Class. Quant. Grav. {\bf 9} (1992) 1487;
    S. Mukohyama and W. Israel, {\it Black Holes, Brick Walls and The
    Boulware State}, Phys. Rev. {\bf D58} (1998) 104005, {\tt [gr-qc/9806012]};
    E. Winstanley, {\it Renormalized Black Hole Entropy via the 'Brick
    Wall' method}, {\tt [hep-th/0011196]}; E. Abdalla and
    L. A. Correa-Borbonet, {\it Black Hole entropy by the Brick Wall
    method in four-dimensions and five-dimensions with $U(1)$
    charges}, {\tt [hep-th/0012101]}.    
\bibitem {ff} V. P. Frolov and D. C. Fursaev, {\it Thermal Fields,
    Entropy, and Black Holes}, Class. Quant. Grav. {\bf 15} (1998)
    2041, {\tt [hep-th/9802010]}.
\bibitem {kkps} S.-W. Kim, W. T. Kim, Y.-J. Park, and H. Shin, {\it
    Entropy of the BTZ Black Hole in (2+1)-Dimensions},
    Phys. Lett. {\bf B392} (1997) 311. 
\bibitem {junction} W. Israel, {\it Singular Hypersurfaces and Thin
    Shells in General Relativity}, Nuov. Ciment. {\bf B44} (1966) 4349.
\bibitem {holo} G. 't Hooft, {\it The Holographic principle:Opening
    Lecture}, {\tt [hep-th/0003004]}; L. Susskind, L. Thorlacius, and
    J. Uglum, {\it The Stretched Horizon and Black Hole
    Complementarity}, Phys. Rev. {\bf D48} (1993) 3743, {\tt
    [hep-th/9306069]}; D. Bigatti and L. Susskind, {\it TASI Lectures
    on the Holographic Principle}, {\tt [hep-th/0002044]}.
\end{references}
\end{document}